\documentclass[11pt]{article}
\usepackage[margin=1in]{geometry}
\usepackage{amsmath,amssymb,amsthm}
\usepackage[colorlinks=true,linkcolor=blue,urlcolor=blue]{hyperref}
\newtheorem{theorem}{Theorem}
\newtheorem{lemma}[theorem]{Lemma}
\newtheorem{proposition}[theorem]{Proposition}
\newcommand{\E}{\mathbb E}
\newcommand{\conv}{\operatorname{conv}}
\newcommand{\diag}{\operatorname{diag}}
\newcommand{\dist}{\operatorname{dist}}
\newcommand{\LH}{\operatorname{LH}}
\newcommand{\clip}{\operatorname{clip}}
\title{Products of Distinguisher Hulls}
\author{John Steinberger \and Zhe Zhang}
\date{November 2, 2026}
\begin{document}
\maketitle
\begin{abstract}
The distinguisher hull of two distributions is the convex hull of the pairs
of acceptance probabilities of tests in a specified class. We study its
behavior under independent products for size-bounded Boolean circuits.
We give a substantially different proof of approximate product containment
for distinguisher hulls, a geometric consequence of Dwork and Tankala's
supersimulator theorem (FOCS 2026). In particular, our construction uses nonadaptive sampling in place of
adaptive score corrections.
\end{abstract}

\section{Introduction}
A distinguishing advantage records one width of a testing region. The entire
region retains the tradeoff between the two acceptance probabilities.
We call its convexified, computationally restricted version a
\emph{distinguisher hull}. With unrestricted tests it is the Lorenz hull of
the two probability measures~\cite{SZ}.

How does this region behave under independent products? For unrestricted
tests, the answer is the coordinate-wise Lorenz product, defined through
products of vector measures~\cite{SZ}. We show that this geometric operation
also controls the computationally restricted regions. Up to an error of
$\eta$ in each acceptance probability, the hull of size-$s$ joint tests is
contained in the product of the individual hulls, with one factor's circuit
budget increased to $O((s+1)\eta^{-2})$ and the other left at $s$.
A reverse inclusion makes points in the geometric product approximately
attainable with larger circuits. Both directions extend to arbitrary finite
products, including repeated sampling from a single pair of distributions.

The forward proof begins with a joint test $D(x,y)$. Fixing $x$ gives a test
of $y$, whose two acceptance probabilities form a point in the second
factor's hull. Sampling $y$ under its two laws approximates this point;
projection returns the approximation to the hull. A probability Lorenz polygon
is a Minkowski sum of nonnegative line segments. Decomposing the projected
point along these segments expresses its coordinates through scalar
coefficient functions of $x$. Their acceptance pairs belong to the first
factor's hull at a larger circuit budget, and the segment decomposition
places the resulting pair in the Lorenz product. Averaging over the sampled
anchors bounds the error. Approximating the polygon by a smaller inscribed
polygon makes the coefficient functions economical to implement.

The sampled sections are selected independently in advance and hardwired
into the circuits; no efficient samplers are assumed. The reverse proof
uses a regularity-based score construction and tests on finite score spaces.

Computational distinguishability under products has been studied by
Halevi and Rabin~\cite{HR}, Geier~\cite{Geier}, Marcussen, Putterman, and
Vadhan~\cite{MPV}. The underlying statistical
geometry appears as privacy regions and trade-off functions in
Kairouz, Oh, and Viswanath~\cite{KOV} and Dong, Roth, and Su~\cite{DRS}.

Recently, Dwork and Tankala~\cite{DT} developed supersimulator machinery
that also yields our forward containment, with matching asymptotic circuit
bounds for both factors. Our proof reaches this conclusion by a different
route: it samples sections of the original distinguisher independently in
advance, whereas the regularity construction selects successive corrections
using the current residual. This distinction concerns the selection of
tests; once selected, the tests in either construction can be evaluated
nonadaptively. Appendix~\ref{app:comparison} gives the deduction from their
supersimulator lemma. Proposition~\ref{prop:optimized} optimizes the
corresponding regularity construction, and the comparison following it in
Appendix~\ref{app:constants} shows that its leading coefficient matches the
limiting leading coefficient of our sampling bound, with different
arithmetic overheads. The contribution here is the geometric formulation
and direct sampling proof. Our reverse attainability argument also
specializes existing regularity and supersimulation methods~\cite{TTV,DT}.

\section{Definitions and forward containment}
Use a fixed universal fan-in-two Boolean circuit basis, with constants and complemented
wires free. Define the size-$s$ distinguisher hull by
\[
 H_s(P,Q)=\conv\{(\E_P f,\E_Q f):f\text{ is a Boolean circuit of size at most }s\}.
\]
For finite input spaces this is a polygon. Constants and complements make it
centrally symmetric about $(1/2,1/2)$, containing $(0,0),(1,1)$ and contained
in $[0,1]^2$. Its lower-boundary edge vectors are nonnegative and sum to
$(1,1)$, so it is a Lorenz hull of a pair of probability measures.
For finitely supported signed measures, write
$\LH(P,Q)=\{(\sum_xP(x)f(x),\sum_xQ(x)f(x)):0\leq f\leq1\}$.
We use the coordinate-wise Lorenz product of Steinberger and Zhang~\cite{SZ},
characterized by
\[
 \LH(P,Q)\odot\LH(P',Q')=\LH(P\otimes P',Q\otimes Q').
\]
Their results establish that the product is independent of the representing
measures, associative, commutative, distributive over Minkowski sums, and
inclusion-preserving. Write $E_\eta=[-\eta,\eta]^2$.

\begin{theorem}\label{thm:main}
There is an absolute constant $C$ such that, for every $s\geq0$, $0<\eta\leq1$,
and distributions $P_X,Q_X$ and $P_Y,Q_Y$ on two finite bit-string spaces,
\begin{equation}\label{eq:main}
 H_s(P_X\otimes P_Y,Q_X\otimes Q_Y)
 \subseteq H_S(P_X,Q_X)\odot H_s(P_Y,Q_Y)+[-\eta,\eta]^2,
 \qquad
 S=\left\lceil C(s+1)\eta^{-2}\right\rceil.
\end{equation}
Consequently, for a pair $P,Q$,
\[
 H_s(P^{\otimes2},Q^{\otimes2})
 \subseteq H_S(P,Q)^{\odot2}+[-\eta,\eta]^2.
\]
\end{theorem}

Only one factor's circuit budget grows in~\eqref{eq:main}. Swapping the two
inputs gives the analogous statement with the other factor enlarged instead.
No efficient sampling algorithm for any of the distributions is assumed.
Sampling below is an averaging argument over hardwired anchors.
For each fixed joint test, all calls to its sections use anchors selected
independently of the input and of the query answers. The postprocessing
may depend nonuniformly on the second factor's hull. This is not an
assertion that the reduction, its anchors, or its geometric data can be
found efficiently.

We first give the core argument with a less economical lookup-table bound,
then prove the geometric implementation that yields the stated size bound.

\section{The geometric factorization}
If a probability Lorenz polygon $L$ has nonnegative segment generators
$v_1,\ldots,v_r$ summing to $(1,1)$, then
\begin{equation}\label{eq:zonotope}
 L=\sum_{j=1}^r[0,v_j],\qquad
 K\odot L=\sum_{j=1}^r\diag(v_j)K.
\end{equation}
Here the sums are Minkowski sums. The second identity follows from
distributivity and the identity $K\odot[0,v]=\diag(v)K$~\cite{SZ}.

\begin{lemma}[Factorization through a polygon]\label{lem:factor}
Let $K$ be a probability Lorenz polygon, let $L$ be as in~\eqref{eq:zonotope},
and let $F$ be a finite-valued statistic
of an input $x$. For each possible value $t$ of $F$, choose coefficients
$\lambda_j(t)\in[0,1]$ and put
\[
 R(t)=\sum_j v_j\lambda_j(t)\in L.
\]
Suppose that, for every $j$, the acceptance pair of
$x\mapsto\lambda_j(F(x))$ belongs to $K$. Then
\[
 \bigl(\E_{P_X}R_1(F(X)),\E_{Q_X}R_2(F(X))\bigr)\in K\odot L.
\]
\end{lemma}
\begin{proof}
The displayed point equals
\[
 \sum_j\diag(v_j)
 \bigl(\E_{P_X}\lambda_j(F(X)),\E_{Q_X}\lambda_j(F(X))\bigr).
\]
Apply~\eqref{eq:zonotope} to its summands.
\end{proof}

The coefficients need not be computed together. Each coefficient may have its
own circuit. This is permitted by the definition of a computational hull and
is the reason the number of generators of $L$ need not enter the first bound.

\section{The core sampling proof}
Fix one size-$s$ Boolean circuit $D(x,y)$ and put $L=H_s(P_Y,Q_Y)$.
For every fixed $x$, its section $y\mapsto D(x,y)$ has size at most $s$, so
\[
 r(x)=\bigl(\E_{P_Y}D(x,Y),\E_{Q_Y}D(x,Y)\bigr)\in L.
\]
Draw anchors $Y_1,\ldots,Y_m$ independently from $P_Y$ and
$Z_1,\ldots,Z_m$ independently from $Q_Y$. Define
\[
 \widehat r(x)=\left(\frac1m\sum_{i=1}^mD(x,Y_i),
                        \frac1m\sum_{i=1}^mD(x,Z_i)\right).
\]
For every fixed $x$,
\begin{equation}\label{eq:variance}
 \E_{\mathrm{anchors}}\|\widehat r(x)-r(x)\|_2^2\leq\frac1{2m}.
\end{equation}
Let $\pi_L$ be Euclidean projection onto $L$. For every point $t$ of the
$(m+1)\times(m+1)$ grid, choose coefficients in $[0,1]$ representing
$\pi_L(t)$ in~\eqref{eq:zonotope}. For fixed anchors and fixed $j$, the function
$x\mapsto\lambda_j(\widehat r(x))$ is a bounded real-valued lookup function
of the two counts.

Every such function is a convex combination of Boolean lookup functions:
\begin{equation}\label{eq:threshold}
 \lambda_j(\widehat r(x))=
 \int_0^1\mathbf1\{\lambda_j(\widehat r(x))\geq u\}\,du.
\end{equation}
There are only finitely many distinct Boolean functions in this integral.
Each uses $2m$ copies of $D$, two population counts, and a lookup on
$(m+1)^2$ possibilities. Its size is $O(m(s+1)+m^2)$.
Thus its acceptance pair, and that of the real-valued coefficient, belong to
$K=H_{O(m(s+1)+m^2)}(P_X,Q_X)$.

Lemma~\ref{lem:factor} shows that for every fixed anchor choice,
\[
 w_{\mathrm{anchors}}=
 \bigl(\E_{P_X}(\pi_L\widehat r)_1,
       \E_{Q_X}(\pi_L\widehat r)_2\bigr)\in K\odot L.
\]
The average $w$ over anchors remains in this convex set. Projection onto a
closed convex set does not increase distance from a point of that set, so
\eqref{eq:variance} and Cauchy--Schwarz give
\[
 \E_{\mathrm{anchors}}\|\pi_L\widehat r(x)-r(x)\|_2\leq\frac1{\sqrt{2m}}.
\]
Integrating its first coordinate under $P_X$ and its second under $Q_X$ yields
\begin{equation}\label{eq:error}
 \left\|w-
  (\E_{P_X\otimes P_Y}D,\E_{Q_X\otimes Q_Y}D)\right\|_\infty
 \leq\frac1{\sqrt{2m}}.
\end{equation}
In a section-query model with arbitrary processing of the answers, this core
proof uses just $2m$ nonadaptive section queries and has error $1/\sqrt{2m}$.
For ordinary circuits, it already proves a direct containment bound with
$S=O((s+1)\eta^{-2}+\eta^{-4})$. Nothing in this argument replaces the original
distributions. The following two elementary geometric observations remove the
large lookup-table term.

\section{Implementing the coefficients economically}
\begin{lemma}[A small inscribed polygon]\label{lem:small}
For a probability Lorenz polygon $L$ and $0<\delta\leq1$, there is a probability
Lorenz polygon $L'\subseteq L$ with $O(1/\delta)$ vertices such that
\[
 \sup_{r\in L}\dist_2(r,L')\leq\delta.
\]
\end{lemma}
\begin{proof}
The lower boundary from $(0,0)$ to $(1,1)$ is coordinate-wise nondecreasing
and has total $\ell_1$ length two. Select its endpoints and additional points
so that successive selected points are at most $\delta$ apart in boundary
$\ell_1$ length. Take the convex hull of these points and their reflections
through $(1/2,1/2)$. This gives $L'\subseteq L$ with $O(1/\delta)$ vertices.
Every point of either boundary chain of $L$ is within Euclidean distance
$\delta$ of a selected point. Every point of $L$ is a convex combination of
boundary points, so the same distance bound holds throughout $L$.
The polygon $L'$ has the required symmetry, endpoints and nonnegative lower
edge vectors, and is therefore a probability Lorenz polygon.
\end{proof}

\begin{lemma}[Coefficient threshold circuits]\label{lem:threshold}
Let $L'$ be a probability Lorenz polygon with $M$ vertices and nonnegative
generators $v_j$. There are functions $\lambda_j:\mathbb R^2\to[0,1]$ such that
\[
 \pi_{L'}(t)=\sum_jv_j\lambda_j(t),
\]
and, on the grid $t=(k/m,\ell/m)$, every threshold
$\mathbf1\{\lambda_j(t)\geq u\}$ is computable from the two binary integers
$k,\ell$ by a Boolean circuit of size $O(M\log^2(m+2))$.
The bound is independent of the precision of the polygon's real coordinates.
\end{lemma}
\begin{proof}
First suppose $L'$ has nonempty interior. At each vertex choose coefficients
in $[0,1]$ representing that vertex, and assign coefficient $1/2$ to every
generator at the center $o=(1/2,1/2)$. Triangulate $L'$ by joining $o$ to its
successive vertices, and interpolate each coefficient affinely on every
triangle. This gives continuous functions on $L'$ with values in $[0,1]$ and
$\sum_jv_j\lambda_j(t)=t$.

Extend these functions by Euclidean projection. Outside the polygon, projection
onto the relative interior of an edge is affine on its outward normal strip;
projection onto a vertex is constant on its outward normal cone. Thus each
extended coefficient is affine on $O(M)$ regions. Each region is the
intersection of at most three affine half-planes: an interior triangle, an
edge strip, or a vertex cone. A coefficient threshold is consequently a union
of $O(M)$ sets, each described by at most four affine half-plane inequalities.

On the finite integer grid $\{0,\ldots,m\}^2$, any affine half-plane dichotomy
can be represented exactly with integer coefficients of magnitude $O(m^2)$.
Here is an elementary justification. Slightly shift the defining threshold
so that it strictly separates its two grid classes, and rescale to give all
points margin at least one. The resulting system has rows
$\pm(k,\ell,1)$ and right-hand side one. Restrict to independent coordinate
directions if its column rank is less than three. The feasible polyhedron then
has no lines and has a vertex. At that vertex, Cramer's rule applied to at most
three independent active constraints gives rational coefficients whose
numerators have magnitude at most $6\max(1,m)^2$. Clearing the determinant
denominator gives integer coefficients with that bound and preserves every
classification. Constant dichotomies need no argument.

Each resulting half-plane test uses arithmetic on $O(\log(m+2))$-bit integers.
Shift-and-add multiplication by hardwired integers and comparison take
$O(\log^2(m+2))$ Boolean gates. Intersect and unite the region predicates to
obtain the claimed bound. Shared boundaries cause no problem because the
coefficient functions agree there.

If $L'$ is degenerate, it is the diagonal segment from $(0,0)$ to $(1,1)$.
Set every coefficient equal to the clipped affine function
$\lambda_j(t)=\min(1,\max(0,(t_1+t_2)/2))$. Since
$\sum_jv_j=(1,1)$, this represents the projection onto the diagonal segment,
and each coefficient threshold satisfies the same bound.
\end{proof}

\section{Proof of the stated circuit bound}
\begin{proof}[Proof of Theorem~\ref{thm:main}]
Repeat the core proof for a fixed $D$, but set
\[
 \delta=\eta/2,\qquad m=\lceil2\eta^{-2}\rceil,
 \qquad L=H_s(P_Y,Q_Y),
\]
and use $L'\subseteq L$ from Lemma~\ref{lem:small}. Replace projection onto $L$
by projection onto $L'$, with the coefficients from
Lemma~\ref{lem:threshold}. Since Euclidean projection is nonexpansive,
\[
 \|\pi_{L'}\widehat r(x)-r(x)\|_2
 \leq\|\widehat r(x)-r(x)\|_2+
       \|\pi_{L'}r(x)-r(x)\|_2.
\]
Taking the anchor expectation bounds this by
$1/\sqrt{2m}+\delta\leq\eta$.

For a fixed anchor choice, each Boolean coefficient-threshold test now has
size
\begin{equation}\label{eq:size}
 2ms+O\bigl(m+\eta^{-1}\log^2(m+2)\bigr)
 =O\bigl((s+1)\eta^{-2}\bigr).
\end{equation}
The counts cost $O(m)$ gates: a balanced addition tree has total size
$O(\sum_{k\geq1} mk/2^k)=O(m)$. The last equality in~\eqref{eq:size} uses
$m=\Theta(\eta^{-2})$ and boundedness of
$\eta\log^2(2+\eta^{-2})$ on $(0,1]$.

Choose the universal constant in $S$ large enough for these circuits.
Equation~\eqref{eq:threshold} puts every coefficient's acceptance pair in
$K=H_S(P_X,Q_X)$. Lemma~\ref{lem:factor}, followed by averaging over anchors,
therefore gives a point in
\[
 K\odot L'\subseteq K\odot L
\]
within $\ell_\infty$ distance $\eta$ of the acceptance pair of $D$.
The inclusion uses the Lorenz product's inclusion preservation. Finally,
$K\odot L+[-\eta,\eta]^2$ is convex, so the same inclusion holds for the convex
hull of all size-$s$ tests $D$.
\end{proof}

\section{Attainability}
\begin{theorem}\label{thm:reverse}
There is a universal constant $C$ such that for all $n\ge2$, $s\ge0$,
and $0<\eta\le1$,
\[
 \bigodot_{i=1}^n H_s(P_i,Q_i)
 \subseteq H_R\left(\bigotimes_{i=1}^nP_i,\bigotimes_{i=1}^nQ_i\right)+E_\eta,
\]
where
\[
 R=\left\lceil C\frac{n^3}{\eta^2}
       \left(s+\log\frac{2n}{\eta}\right)\right\rceil.
\]
In particular, $H_s(P,Q)^{\odot n}\subseteq
H_R(P^{\otimes n},Q^{\otimes n})+E_\eta$.
\end{theorem}

Attainability here is in the convex hull of circuit acceptance pairs.
It does not mean that every interior point is attained by one deterministic
circuit. All constructions are nonuniform existence arguments; neither
efficient samplers nor algorithms for finding the circuits are assumed.

\section{A score whose hull approximately dominates the given hull}
The construction below is a specialization of the supersimulator method
of Dwork and Tankala~\cite[Lemma 2.6]{DT}, in the regularity tradition
of Trevisan, Tulsiani, and Vadhan~\cite{TTV}. We include the short
potential argument and its translation into hull containment.
\begin{lemma}\label{lem:score}
For each integer $m\ge1$ and pair $P,Q$, there is a circuit computing a score
$p(x)\in\{0,1/m,\ldots,1\}$, encoded by its integer numerator, of size
\[
 T=O\bigl(m^2(s+\log(m+2))\bigr),
\]
such that, writing $L=\LH(p_*P,p_*Q)$ for the ordinary Lorenz hull of its
two output laws,
\[
 H_s(P,Q)\subseteq L+E_{4/m}.
\]
\end{lemma}
\begin{proof}
Let $\mu=(P+Q)/2$ and $\rho(x)=Q(x)/(P(x)+Q(x))$ on the support of $\mu$.
Set $\rho=0$ elsewhere. Then $dQ=2\rho\,d\mu$ and
$dP=2(1-\rho)\,d\mu$. Put $\varepsilon=1/m$ and start with $p=0$.

For the current score, consider every size-$s$ Boolean test and every
threshold test $\mathbf1\{p\ge t\}$. If any such test $f$ satisfies
\[
 |\E_\mu[(\rho-p)f]|>\varepsilon,
\]
choose one, let $\sigma$ be the sign of this expectation, and replace $p$ by
\[
 p'=\clip_{[0,1]}(p+\sigma\varepsilon f).
\]
The potential $\Phi(p)=\E_\mu(\rho-p)^2$ obeys
\[
 \Phi(p')\le\Phi(p)-2\varepsilon
        |\E_\mu[(\rho-p)f]|+\varepsilon^2\E_\mu f
 <\Phi(p)-\varepsilon^2.
\]
Clipping can only improve the squared distance to $\rho\in[0,1]$.
Since $0\le\Phi\le1$, fewer than $m^2$ updates are possible. At termination,
\begin{equation}\label{eq:residual}
 |\E_\mu[(\rho-p)f]|\le\varepsilon
\end{equation}
holds both for all size-$s$ tests and for every threshold of the final score.

The score remains on the stated grid. An update appends either a size-$s$
circuit or a comparison of the current integer score with a hardwired
integer, followed by a conditional increment or decrement and saturation.
It costs $O(s+\log(m+2))$ gates. Earlier score circuits are shared, not
copied. This proves the size bound, including for updates that use a threshold
of the current score. Finding a violating test is not asserted to be efficient.

It remains to translate~\eqref{eq:residual} into hull domination.
Normalize a mixed-sign support direction to $(-a,b)$, with $a,b\ge0$ and
$a+b=1$. For every size-$s$ test $f$,
\begin{align*}
 -a\E_P f+b\E_Q f
 &=2\E_\mu[(\rho-a)f]\\
 &\le2\E_\mu[(p-a)_+]+2\varepsilon.
\end{align*}
The threshold $g=\mathbf1\{p\ge a\}$ has acceptance pair in $L$, and
\[
 -a\E_P g+b\E_Q g
 =2\E_\mu[(p-a)_+]+2\E_\mu[(\rho-p)g]
 \ge2\E_\mu[(p-a)_+]-2\varepsilon.
\]
Thus the support functions satisfy
$h_{H_s}(-a,b)\le h_L(-a,b)+4\varepsilon$.
Both hulls are centrally symmetric about $(1/2,1/2)$, so the same bound
holds in the opposite mixed-sign directions. In same-sign directions their
support functions agree, since both are in $[0,1]^2$ and contain $(0,0),(1,1)$.
By homogeneity the error in any direction $u$ is at most
$4\varepsilon\|u\|_1=h_{E_{4\varepsilon}}(u)$, proving the inclusion.
\end{proof}

Only residual correlations with Boolean tests and score thresholds are used.
There is no need here to control the residual separately on every score
level, or to prove indistinguishability of replacement distributions.

\section{Two elementary product facts}
\begin{lemma}\label{lem:stability}
If $K_i,L_i$ are probability Lorenz hulls and
$K_i\subseteq L_i+E_{\delta_i}$, then
\[
 \bigodot_iK_i\subseteq\bigodot_iL_i+E_{\sum_i\delta_i}.
\]
\end{lemma}
\begin{proof}
For every probability Lorenz hull $A$, $A\odot E_\delta=E_\delta$.
Indeed, $E_\delta$ is the sum of the four axis segments generated by
$\delta e_1,-\delta e_1,\delta e_2,-\delta e_2$. The projection of $A$ on
either coordinate axis is $[0,1]$. Distributivity and the segment product
identity give the assertion. Replace the factors one at a time, using
inclusion preservation and this identity. Every product of the other,
unreplaced or replaced, factors is a probability Lorenz hull.
The errors therefore add. No product of two error boxes is needed.
\end{proof}

\begin{lemma}\label{lem:finite}
Suppose $F_i$ are deterministic statistics taking values in
$\{0,\ldots,M-1\}$, each computable by a circuit of size at most $T$,
with outputs encoded as binary indices.
Let $L_i=\LH((F_i)_*P_i,(F_i)_*Q_i)$. For $0<\gamma\le1$,
\[
 \bigodot_iL_i\subseteq
 H_U\left(\bigotimes_iP_i,\bigotimes_iQ_i\right)+E_\gamma,
\]
where, for $N=\lceil2nM/\gamma\rceil$,
\[
 U=O\bigl(nT+nM\log(MN+2)+n^2\log^2(N+2)\bigr).
\]
\end{lemma}
\begin{proof}
Round each of the $2n$ output probability vectors to a probability vector
with denominator $N$. Taking floors and distributing the missing units gives
total variation error at most $M/N\le\gamma/(2n)$ in each vector.
Each of the two product laws consequently changes by at most $\gamma/2$
in total variation.

For the rounded laws, the product hull is a finite zonotope. Every vertex
is attained by a likelihood-ratio threshold test, with constants and
complements included. On the score tuple $z=(z_1,\ldots,z_n)$, such a test
compares the two products of hardwired integer probability numerators
\[
 A(z)=\prod_i a_i(z_i),\qquad B(z)=\prod_i b_i(z_i),
 \qquad 0\le a_i(z_i),b_i(z_i)\le N.
\]
The relevant likelihood ratios are ratios of integers at most $N^n$.
Any threshold cut between distinct ratios can therefore be specified by a
rational threshold with numerator and denominator of $O(n\log(N+2))$ bits
(use a mediant between consecutive ratios). Endpoint cuts, zero probabilities,
and infinite ratios are handled by zero tests and constants. At a supporting
edge, including or excluding an entire tied ratio class gives its endpoints;
convexification supplies the edge.

Each lookup of $a_i,b_i$ costs $O(M\log(MN+2))$ gates, by decoding the
score and selecting hardwired bits. Balanced schoolbook multiplication
computes the two products with $O(n^2\log^2(N+2))$ gates; the final
integer comparison, including multiplication by threshold constants, has
the same bound. Adding the statistic circuits gives $U$.

To approximate an arbitrary point of the actual product hull, use its
acceptance function under the rounded laws first. Its two coordinates change
by at most $\gamma/2$. Express that rounded acceptance pair as a convex
combination of vertices of the rounded hull, and implement their threshold
circuits as above. Evaluating this same convex combination under the actual
laws changes the coordinates by at most another $\gamma/2$. Its acceptance
pair is in $H_U$, proving the assertion.
\end{proof}

\section{Products of arbitrarily many factors}
\begin{proof}[Proof of Theorem~\ref{thm:reverse}]
Set $m=\lceil8n/\eta\rceil$. Apply Lemma~\ref{lem:score} separately to
each factor. Lemma~\ref{lem:stability} gives
\[
 \bigodot_iH_s(P_i,Q_i)\subseteq\bigodot_iL_i+E_{4n/m}
 \subseteq\bigodot_iL_i+E_{\eta/2}.
\]
Apply Lemma~\ref{lem:finite} with $M=m+1$, $\gamma=\eta/2$, and
$T=O(m^2(s+\log(m+2)))$. Then $N=O(n^2/\eta^2)$ and its circuit bound is
\[
 O\left(\frac{n^3}{\eta^2}\left(s+\log\frac{2n}{\eta}\right)
       +\frac{n^2}{\eta}\log\frac{2n}{\eta}
       +n^2\log^2\frac{2n}{\eta}\right).
\]
The first term absorbs the others for $n\ge2$, $0<\eta\le1$.
The two error boxes sum to $E_\eta$.
\end{proof}

Together with Theorem~\ref{thm:main}, this gives the following
two directions, at different circuit budgets:
\begin{align*}
 H_s(P^{\otimes n},Q^{\otimes n})
 &\subseteq H_S(P,Q)^{\odot n}+E_\eta,
 & S&=O((s+1)n^2/\eta^2),\\
 H_s(P,Q)^{\odot n}
 &\subseteq H_R(P^{\otimes n},Q^{\otimes n})+E_\eta,
 & R&=O\left(n^3\eta^{-2}(s+\log(2n/\eta))\right).
\end{align*}
For the first inclusion, iterate Theorem~\ref{thm:main} with
error $\eta/(n-1)$, leaving the remaining block at budget $s$ at every step.
The identity used in Lemma~\ref{lem:stability} makes these errors additive.
This is not a same-budget Hausdorff approximation or a claim of sharp overhead.

\appendix
\section{Relation to prior results}\label{app:comparison}
This appendix records implications between specific statements. A different
formulation or proof does not by itself establish a strict strengthening or
incomparability. In particular, the hull-valued nature of our conclusions
alone does not separate them from the proxy constructions below.

\paragraph{Testing regions and trade-off functions.}
For unrestricted tests, write
\[
 H(P,Q)=\{(\E_P f,\E_Q f):0\le f\le1\}.
\]
For finite distributions this equals the convex hull of the acceptance pairs
of Boolean tests. In type-I/type-II error coordinates, the corresponding
region is the image of $H(P,Q)$ under $(x,y)\mapsto(x,1-y)$.
Kairouz--Oh--Viswanath call the region for a fixed pair of output distributions
a privacy region~\cite[Eq.~(3)]{KOV}. Dong--Roth--Su instead use its optimal
boundary, the trade-off function
\[
 T(P,Q)(\alpha)=1-\sup\{y:(x,y)\in H(P,Q),\ x\le\alpha\}.
\]
Their Appendix A also discusses the equivalent testing-region representation,
restricted to tests at least as good as random guessing. Central symmetry
recovers the other half of the full region.
Their product is defined by
\[
 T(P,Q)\otimes T(P',Q')=T(P\otimes P',Q\otimes Q').
\]
They prove that this product is well-defined, associative and monotone
\cite[Definition 3.1 and Appendix D]{DRS}. For pairs of probability measures,
it represents the coordinate-wise Lorenz product through its boundary. The statistical product geometry is prior work. The circuit bounds
in the present paper concern its computational restriction.

\paragraph{An asymmetric hull consequence of MPV.}
We use the corrected version of the partition theorem in
\cite[Theorem 3.1]{MPV}. Given $P,Q$ and the size-$s$ test class, it supplies
proxies $\widetilde P,\widetilde Q$ and a statistic $p$ taking $M$ values,
such that each original is $\varepsilon$-indistinguishable from its proxy,
\[
 p_*P=p_*\widetilde P,\qquad p_*Q=p_*\widetilde Q,
\]
and the two proxies have the same conditional distribution $\nu_j$ on
each cell $p=j$. Null cells can be assigned arbitrary conditional laws.
Set $L=\LH(p_*P,p_*Q)$ and
$v_j=(P[p=j],Q[p=j])$, so that $L=\sum_j[0,v_j]$.

Fix a size-$s$ joint test $D(x,y)$. Replacing its first input by the respective
proxy changes each acceptance probability by at most $\varepsilon$:
fixing $y$ leaves a size-$s$ test of $x$, and one then averages over $y$.
For each cell let $g_j(y)=\E_{x\sim\nu_j}D(x,y)$. Its acceptance pair under
$P_Y,Q_Y$ belongs to $H_s(P_Y,Q_Y)$, since it is a convex combination of
acceptance pairs of size-$s$ sections. The proxy joint acceptance pair is
\[
 \sum_j\diag(v_j)(\E_{P_Y}g_j,\E_{Q_Y}g_j)
 \in L\odot H_s(P_Y,Q_Y).
\]
If computing $p$ costs $T_p$ gates, every Boolean function of $p$ costs
$T_p+O(M\log(M+2))$ gates by a finite lookup. With this budget denoted by
$T$, $L\subseteq H_T(P,Q)$, giving
\begin{equation}\label{eq:mpv-forward}
 H_s(P\otimes P_Y,Q\otimes Q_Y)
 \subseteq H_T(P,Q)\odot H_s(P_Y,Q_Y)+E_\varepsilon.
\end{equation}
The right-hand side is convex, so the argument extends from individual tests
to the entire hull. This proves that both the shape conclusion and the
asymmetric form follow from MPV's partition theorem, with its resource cost.

For input length $\ell=\log_2|\mathcal X|$ on a full bit-string domain,
the corrected bound in that theorem gives
\[
 T_p=O\bigl(s\varepsilon^{-6}
       +\varepsilon^{-12}\ell\log(|\mathcal X|/\varepsilon)\bigr),
 \qquad M=O(\varepsilon^{-1}).
\]
Theorem~\ref{thm:main} gives the smaller bound
$O((s+1)\varepsilon^{-2})$ for the same containment. This comparison is
with the explicit derivation~\eqref{eq:mpv-forward}; it is not a claim that
all of MPV's conclusions follow with improved parameters. Their theorem
also constructs proxies with properties not supplied by our forward proof.

The same partition theorem yields qualitative attainability. Computational
closeness and the common conditional laws imply
$H_s(P,Q)\subseteq L+E_\varepsilon$. Lemma~\ref{lem:stability} transfers
this inclusion to products. The statistic has a finite range, so
Lemma~\ref{lem:finite} implements the product hull with polynomial overhead.
Hence polynomial-overhead attainability itself is also a consequence of the
partition construction, even though its bounds differ from ours.

\paragraph{The reverse bound from Dwork--Tankala.}
The score step of our proof follows from the general supersimulator
lemma~\cite[Lemma 2.6]{DT}. With
$\mu=(P+Q)/2$ and $g=Q/(P+Q)$, choose the growing test family
\[
 \mathcal G(h)=\mathcal C_s\cup
 \{\mathbf1\{h\ge t\}:t\in[0,1]\},
\]
where $\mathcal C_s$ denotes the size-$s$ Boolean tests.
The lemma supplies $h$ whose residual correlations with every member of
$\mathcal G(h)$ are at most $\varepsilon$.
These are precisely the two conditions used in Lemma~\ref{lem:score}.

For reciprocal-integer error, their construction uses
$O(\varepsilon^{-2})$ signed updates with a fixed step of order
$\varepsilon$. The running sum can be stored as an integer with
$O(\log(2/\varepsilon))$ bits, and its clipped value has
$O(1/\varepsilon)$ possible outputs. At each step the added test is either
a size-$s$ circuit or a threshold of the existing score. Sharing that score
circuit gives total size
\[
 O\bigl(\varepsilon^{-2}(s+\log(2/\varepsilon))\bigr).
\]
Their use of clipping the accumulated sum, rather than clipping after every
update as above, does not change this circuit accounting or the final
residual guarantee. The support-function argument in Lemma~\ref{lem:score}
and Lemmas~\ref{lem:stability}--\ref{lem:finite} then give
Theorem~\ref{thm:reverse} with the stated order of circuit size.
Thus the reverse theorem is a geometric application of their machinery;
the score construction is not an independent strengthening of it.

\paragraph{The forward bound from Dwork--Tankala.}
Theorem~\ref{thm:main}, including its asymmetric circuit budgets, also follows
from their supersimulator lemma. We give the derivation since it is not
stated in this geometric form in~\cite{DT}.
Write $P=P_X$, $Q=Q_X$, $\mu=(P+Q)/2$, and $g=Q/(P+Q)$, assigning
an arbitrary value where the denominator vanishes. Apply the lemma with
the growing family consisting of all size-$s$ tests, all thresholds of $h$,
and their complements. For $0<\delta\leq1/8$, the resulting score satisfies
\[
 |\E_\mu f(g-h)|\leq\delta
\]
for every such test. The shared-counter implementation described above
computes each threshold and its complement with size
\[
 T=O\bigl(\delta^{-2}(s+\log(2/\delta))\bigr).
\]
One can take reciprocal-integer $\delta$ within a constant factor of any
specified tolerance. Since the constant test is included,
$|\E_\mu h-1/2|\leq\delta$. Define probability measures
\[
 \widetilde P=\frac{(1-h)\mu}{\E_\mu(1-h)},\qquad
 \widetilde Q=\frac{h\mu}{\E_\mu h}.
\]
For every test in the family, replacing $P,Q$ by these proxies changes
each acceptance probability by at most $4\delta$: replacing $g$ by $h$
costs $2\delta$, and normalizing costs at most another $2\delta$.

The proxy likelihood ratio is monotone in $h$. Their statistical hull
$A=\LH(\widetilde P,\widetilde Q)$ is therefore the convex hull of the
acceptance pairs of score thresholds, their complements, and constants.
Consequently
\[
 A\subseteq H_T(P,Q)+E_{4\delta}.
\]
Now fix a size-$s$ joint test $D(x,y)$ and put $L=H_s(P_Y,Q_Y)$.
Conditional on any score value of positive $\mu$-mass, the proxies have
the common conditional law $\mu(\,\cdot\mid h)$ whenever that conditioning
event has positive proxy mass. Averaging $D(x,y)$ over this law gives a
convex combination of size-$s$ tests of $y$, hence an acceptance pair in $L$.
Decomposing $A$ into the segments given by its score masses shows that the
proxy joint acceptance pair lies in $A\odot L$.
The original joint pair differs from it by at most $4\delta$ in each
coordinate, since every section $D(\,\cdot,y)$ has size at most $s$.
Lemma~\ref{lem:stability} now gives
\[
 H_s(P_X\otimes P_Y,Q_X\otimes Q_Y)
 \subseteq H_T(P_X,Q_X)\odot H_s(P_Y,Q_Y)+E_{8\delta}.
\]

It remains to remove the apparent logarithmic overhead. Take
$\delta$ of order $\eta/8$, with $8\delta\leq\eta$.
If $s\geq\log_2(1/\eta)$, then
$T=O((s+1)\eta^{-2})$ already. Otherwise, fix $D$ again and discard gates
not connected to its output. A fan-in-two circuit with at most $s$ gates
reads at most $s+1$ distinct input bits, so it reads at most $s+1$ bits of
$x$. Let $F(x)$ record those bits. The section acceptance pair $r(x)\in L$
depends only on $F(x)$. Choose a segment decomposition of $r(x)$ separately
for each value of $F(x)$. Each coefficient threshold is a Boolean function
of at most $s+1$ bits and has a truth-table circuit of size
$O((s+1)2^{s+1})$. Lemma~\ref{lem:factor} therefore gives exact containment
for this joint test with that first-factor budget. Since
$2^s<1/\eta$, this budget is at most $O((s+1)\eta^{-2})$.
Taking convex hulls completes the derivation of Theorem~\ref{thm:main}
with its stated bound in both regimes.

\paragraph{Scalar product bounds.}
Geier's two-factor theorem bounds computational distinguishing advantage by
$d_X+d_Y-d_Xd_Y$ plus an error decreasing exponentially with a reduction
parameter, when $d_Y<1$ is fixed~\cite[Theorem 3.1]{Geier}. The corresponding
circuit overhead has logarithmic dependence on inverse error, apart from
input-length costs. The quadratic inverse-error dependence of our forward
bound does not recover that efficiency. Geier also treats uniform reductions,
whereas our statements concern nonuniform circuits.

The hull retains information absent from these two scalar inputs. For example,
the probability pairs
\[
 ((3/4,1/4),(1/4,3/4))\quad\hbox{and}\quad((1,0),(1/2,1/2))
\]
both have total variation distance $1/2$, but their squares have total
variation distances $1/2$ and $3/4$, respectively. Thus the same scalar
product bound applies to both while their hull products differ.
This certifies loss of information in the scalar summary, not a theorem
that no further use of Geier's methods can yield shape-sensitive bounds.
Halevi--Rabin~\cite{HR} additionally study sequential interactive protocols;
our independent-product statements do not address that setting.

Fehr--Vaudenay~\cite{FV} obtain statistical repeated-sampling inequalities
under additional information about the distributions. Such inequalities
can be applied to the probability experiments representing the hull factors
in our forward theorem. They provide bounds on the product's thickness,
with at most $2\eta$ additional distinguishing advantage from an
$E_\eta$ error box. This is a way to combine the results, rather than an
incomparability claim.

\section{Structure of the reductions and explicit constants}\label{app:constants}
\paragraph{Selection and evaluation.}
For a fixed joint test $D$, the sampling proof draws anchors independently
from $P_Y$ and $Q_Y$. Each coefficient circuit queries the same fixed list
of sections $D(x,Y_i),D(x,Z_i)$ and then processes their answers.
The queries are nonadaptive, and their selection uses no residual tests.
The construction remains nonuniform: the hull geometry is supplied as
hardwired data, and averaging over anchors establishes containment in a
convex hull rather than a deterministic efficient procedure for choosing
anchors.

The regularity proof below selects a test using its correlation with the
current residual and updates a score. Both the selection and the score
construction are sequential. This does not imply adaptive oracle queries
during evaluation of the final circuit. Once the selected small circuits
are fixed, they can be evaluated in parallel; their outputs feed the
successive score updates. Moreover, for a fixed $D$ one can restrict the
small-test family to its sections $D(\,\cdot,y)$: the proof below uses only
convex combinations of those sections. This version also has a fixed list
of section queries at evaluation time, although it selects that list by
residual-dependent corrections rather than independent sampling.
The structural distinction is therefore independent sampling versus
residual-dependent selection, not a proved separation between adaptive
and nonadaptive oracle reductions. No lower bound on the parallel depth
of the regularity implementation is asserted.

\paragraph{A representation independent of the target.}
The sampling argument also gives a restricted simultaneous simulation
statement. Fix an input law $\mu$ and the two-coordinate profile $r$
from the sampling proof. For $0<\varepsilon\le1$,
choose $m=O(\varepsilon^{-2})$ anchors per coordinate and round the
empirical profile to a grid of mesh $O(\varepsilon)$. Some fixed anchor
choice gives a statistic $T$ with $O(\varepsilon^{-2})$ values and
$\|r-T\|_{L^2(\mu)}\le\varepsilon$, by the variance bound and rounding.
This choice is independent of any target $g:X\to[0,1]$.
For each such target put $h=\E_\mu[g\mid T]$. Simultaneously for all
$\Phi:[0,1]^2\times[0,1]\to[-1,1]$ that are $1$-Lipschitz in the first
argument, uniformly in the second,
\[
 \bigl|\E_\mu[(g-h)\Phi(r,h)]\bigr|
 =\bigl|\E_\mu[(g-h)(\Phi(r,h)-\Phi(T,h))]\bigr|
 \le \tfrac12\varepsilon.
\]
Indeed, the subtracted term has expectation zero conditional on $T$,
and $\|g-h\|_2\le1/2$. No regularity in the score argument is required.
Rounding $h$ to a $T$-measurable score $\widetilde h$ within
$\varepsilon/2$ gives the bound $\varepsilon$ with $\widetilde h$ in
both occurrences: the same orthogonality applies to $\Phi(T,\widetilde h)$.
Thus one list of nonadaptive section queries serves every target, with a
different conditional-expectation table for each. These tables are
nonuniform data, not efficiently learned here. The argument is an elementary
discretization and conditional-expectation principle, akin to partitioning
regression~\cite{Nobel}; it uses no special Lorenz geometry and gives no
improvement over the general supersimulator theorem for unrestricted
test families.

\paragraph{Printed constants and constants of the deduction.}
Dwork--Tankala's Lemmas 2.4 and 2.6 print the update bound $2/\delta^2$,
with step size $\delta/2$~\cite{DT}. Their main product theorem gives
circuit costs in asymptotic notation. Combining the printed update bound
with the normalized-proxy deduction in Appendix~\ref{app:comparison},
whose error is $8\delta$, gives $128/\eta^2$ updates at $\delta=\eta/8$.
The number $128$ belongs to this deduction; it is not a bound printed in
their paper for a hull theorem.

Two elementary tightenings reduce it. Initializing at $h_0=1/2$ bounds
$\E_\mu(g-h_0)^2$ by $1/4$. Their prefix-sum proof then gives
$1/(2\delta^2)$ updates, without changing the construction. This sharpens
the estimate for this initialization, not their statement for an arbitrary
prescribed $h_0$. Next, normalization of the proxies is unnecessary:
use $\widehat P=2(1-h)\mu$ and $\widehat Q=2h\mu$. Each tested expectation
changes by at most $2\delta$, and the proxy hull is within $E_{2\delta}$
of the computational hull by the threshold argument. Although these
measures need not be probabilities, multiplication by the unchanged
probability hull still preserves the error box, by distributivity and
$E_\epsilon\odot L=E_\epsilon$. The total error is therefore $4\delta$.
Together these observations give $8/\eta^2$ updates.

Projecting after every update, instead of clipping the accumulated sum,
gives a further factor of two: with step $\delta$, a violated residual test
reduces squared error by more than $\delta^2$. This gives
$1/(4\delta^2)$ updates from initialization $1/2$, or $4/\eta^2$ for
error $4\delta$. This modification preserves the residual guarantee but
changes the structured-sum representation specified in their lemma.
The following specialization improves the error accounting as well.
Its numerical bound is derived here from their regularity method; it is
not an explicitly stated bound in~\cite{DT}.

\begin{proposition}\label{prop:optimized}
For $0<\eta\leq1$, the asymmetric containment
\[
 H_s(P_X\otimes P_Y,Q_X\otimes Q_Y)
 \subseteq H_S(P_X,Q_X)\odot H_s(P_Y,Q_Y)+E_\eta
\]
holds with
\[
 S\leq \lceil\eta^{-2}\rceil s
       +O\bigl(\lceil\eta^{-2}\rceil\log(2/\eta)\bigr).
\]
More precisely, the construction uses $N<\kappa/\eta^2$ updates when
$\kappa>0$, each containing at most one size-$s$ circuit, where
\[
 \kappa=\frac12\sum_x
 \frac{(P_X(x)-Q_X(x))^2}{P_X(x)+Q_X(x)}\leq1.
\]
Terms with zero denominator are zero. When $\kappa=0$, no updates are needed.
The remaining circuit cost is $O((N+1)\log(2/\eta))$.
\end{proposition}

\begin{proof}
Put $P=P_X$, $Q=Q_X$, $\mu=(P+Q)/2$, and $g=Q/(P+Q)$ on the
support of $\mu$. Set $\delta=\eta/2$ and initialize $h_0=1/2$.
For each current score $h$, consider the growing family
\[
 \mathcal G(h)=\{f-t:\ \operatorname{size}(f)\leq s,
       \ t=\mathbf1\{h\geq a\},\ a\in\mathbb R\}.
\]
This includes constant thresholds. Every member has absolute value at
most one. If $u\in\mathcal G(h)$ violates
$|\E_\mu(g-h)u|\leq\delta$, choose its correlation sign $\sigma$ and set
\[
 h'=\clip_{[0,1]}(h+\sigma\delta u).
\]
The potential satisfies
\begin{align*}
 \E_\mu(g-h')^2
 &\leq\E_\mu(g-h)^2-2\delta|\E_\mu(g-h)u|
                      +\delta^2\E_\mu u^2\\
 &<\E_\mu(g-h)^2-\delta^2.
\end{align*}
Initially the potential is
$\E_\mu(g-1/2)^2=\kappa/4\leq1/4$.
Thus the number of updates is strictly less than $\kappa/(4\delta^2)$
if positive, and the final score satisfies
\begin{equation}\label{eq:optimized-diff}
 |\E_\mu(g-h)(f-t)|\leq\delta
 \quad\text{for all small tests $f$ and thresholds $t$ of $h$.}
\end{equation}
The same estimate holds if $f$ and $t$ are independently replaced by convex
combinations of their respective classes.

Each update evaluates one size-$s$ test and a threshold of the preceding
score, then conditionally increments, decrements, or leaves the score
unchanged, with clipping. The preceding score circuit is shared.
There is no precision assumption on $\eta$: every score value is of the
form $b+k\delta$, with $b\in\{0,1/2,1\}$ and an integer
$|k|=O(1/\delta)$. Store the choice of $b$ and the integer $k$.
Clipping resets this representation to an endpoint. Threshold comparisons
use hardwired integer cutoffs for each of the three choices of $b$.
Consequently each update costs at most $s+O(\log(2/\eta))$ gates,
and a final threshold costs $O(\log(2/\eta))$ gates.

To prove containment, fix a size-$s$ joint test $D(x,y)$ and set
\[
 r_P(x)=\E_{P_Y}D(x,Y),\qquad
 r_Q(x)=\E_{Q_Y}D(x,Y),\qquad L=H_s(P_Y,Q_Y).
\]
Every $(r_P(x),r_Q(x))$ lies in $L$, and each scalar function $r_P,r_Q$
is a convex combination of size-$s$ tests of $x$.
Consider a support direction $(-a,b)$ with $a,b\geq0$, $a+b=1$.
Write $L=\sum_j[0,(p_j,q_j)]$ as above, discarding zero generators.
For $z\in[0,1]$ choose
\[
 t_j(z)=\mathbf1\{-a(1-z)p_j+bzq_j\geq0\},\qquad
 U(z)=\sum_jp_jt_j(z),\quad V(z)=\sum_jq_jt_j(z).
\]
Each $t_j$ is a threshold, or a constant, and $(U(z),V(z))$ maximizes
$-a(1-z)u+bzv$ over $(u,v)\in L$. In particular, at $z=h(x)$,
\[
 -a(1-h)(r_P-U(h))+bh(r_Q-V(h))\leq0.
\]
Writing $K=H_S(P,Q)$, the point
\[
 w=(\E_P U(h),\E_Q V(h))
   =\sum_j\operatorname{diag}(p_j,q_j)
       (\E_P t_j(h),\E_Q t_j(h))
\]
belongs to $K\odot L$, since each threshold of $h$ has size at most $S$.
Let $z_D=(\E_{P\otimes P_Y}D,\E_{Q\otimes Q_Y}D)$.
Subtracting the two support evaluations yields
\begin{align*}
 \langle(-a,b),z_D-w\rangle
 &=2\E_\mu\bigl[-a(1-g)(r_P-U(h))+bg(r_Q-V(h))\bigr]\\
 &\leq2\E_\mu(g-h)
             \bigl[a(r_P-U(h))+b(r_Q-V(h))\bigr]\\
 &\leq2\delta=\eta.
\end{align*}
The last step uses~\eqref{eq:optimized-diff}: $U(h)$ and $V(h)$ are convex
combinations of score thresholds, with weights summing to one.
Taking the supremum over $D$ proves the support inequality in this quadrant.
Both hulls are centrally symmetric about $(1/2,1/2)$, which gives the opposite
mixed-sign quadrant. In same-sign directions the support functions agree,
since both hulls lie in $[0,1]^2$ and contain $(0,0),(1,1)$.
Homogeneity now gives the error $\eta\|v\|_1$ in every direction $v$,
which is precisely containment up to $E_\eta$.
\end{proof}

\paragraph{Comparison with sampling.}
Give a fraction $\theta\in(0,1)$ of the sampling proof's error budget to
polygon approximation. Its explicit cost is
\[
 2\left\lceil\frac{1}{2(1-\theta)^2\eta^2}\right\rceil s
 +O\left(\frac{1}{(1-\theta)^2\eta^2}
 +\frac{\log^2(2+((1-\theta)\eta)^{-2})}{\theta\eta}\right).
\]
For fixed $\theta$, its coefficient of $s/\eta^2$ is $(1-\theta)^{-2}$,
approaching one as $\theta$ decreases. The optimized regularity proof
already has coefficient one (up to integer rounding), but its displayed
arithmetic overhead is $O(\eta^{-2}\log(2/\eta))$ rather than
$O_\theta(\eta^{-2})$. These bounds therefore do not establish a uniform
total-gate advantage in either direction. They also do not prove that
coefficient one is optimal for either approach.

The apparent coefficients $128$, $8$, and $4$ in successively tightened
proxy-based derivations are not barriers. The change to differences of
tests avoids estimating two residual correlations separately and reduces
the final support error from $4\delta$ to $2\delta$; the update count falls
by a factor of four. Neither proxy normalization nor proxy measures are
needed in this optimized specialization.

For reference, the successive bounds on the number of updates, each
containing at most one size-$s$ circuit, are
\[
\begin{array}{l|c|c}
\text{Derivation} & \text{Hull error} & \text{Update bound at error }\eta\\ \hline
\text{Printed lemma and normalized-proxy deduction} & 8\delta & 128\eta^{-2}\\
\text{Initialization }1/2\text{ and unnormalized proxies} & 4\delta & 8\eta^{-2}\\
\text{Projection after every update} & 4\delta & 4\eta^{-2}\\
\text{Differences of tests, Proposition~\ref{prop:optimized}} & 2\delta & \eta^{-2}
\end{array}
\]
These are bounds from particular derivations, excluding arithmetic overhead,
not necessary costs or a constant-factor separation from the published
results. The small-circuit truth-table argument in
Appendix~\ref{app:comparison} still recovers the uniform
$O((s+1)\eta^{-2})$ bound when the logarithmic arithmetic term would dominate.

\end{document}